\documentstyle[preprint,eqsecnum,aps]{revtex}
\begin{document}
\def\ie{{\it i.e.\ }}
\def\R{\mbox{\bf R}}
\def\paren#1{\left( #1 \right)}

\def\crit{critical}
\def\superc{supercritical}
\def\subc{subcritical}
\def\bh{black hole}
\def\bhs{black holes}
\def\MB{M_B}
\def\xB{x_B}
\def\xI{x_I}
\def\xcut{x_{\rm cut}}
\def\ycut{y_{\rm cut}}
\def\DyI{\Delta y_I}
\def\yI{y_I}
\def\tM{\tilde M}
\def\tMT{\tilde M_T}
\def\tMC{\tilde M_C}
\def\tMB{\tilde M_B}
\def\tMI{\tilde M_I}
\def\DtMI{\Delta\tilde M_I}
\def\tR{\tilde R}
\def\tRS{\tilde R_S}
\def\tRI{\tilde R_I}
\def\tK{\tilde K}
\def\dR{{\dot{R}}}
\def\dM{{\dot{M}}}
\def\xmin{x_{\rm min}}
\def\xmax{x_{\rm max}}

\preprint{TIT/HEP-266/COSMO-46}
\title{An analytic model with critical behavior in black hole
  formation} 
\author{Tatsuhiko Koike\thanks{JSPS fellow, e-mail:
    bartok@phys.titech.ac.jp}}
\address{Department of Physics, Tokyo Institute of Technology,
Oh-Okayama, Meguro, Tokyo 152, Japan}
\author{\em and}
\author{Takashi Mishima\thanks{e-mail:
    tmishima@phys.titech.ac.jp}} 
\address{Laboratories of Physics, College of Science and Technology, 
Nihon University, Narashinodai, Funabashi, Chiba 274, Japan}
\date{September, 1994}
\maketitle
\begin{abstract}
A simple analytic model is presented which exhibits a critical
behavior in  black hole formation, namely, collapse of a thin shell
coupled with outgoing null fluid. It is seen that the critical
behavior is caused by the gravitational nonlinearity near the event
horizon.  We calculate the value of the critical exponent analytically
and find that it is very dependent on the coupling constants of the
system. 
\end{abstract}

\section{Introduction}
\label{intro}
Gravitational collapse with formation of black holes is one of the  
main problems of general relativity. Choptuik \cite{Cho93} discovered
a critical behavior of the gravitational collapse of massless scalar
field by a numerical study. His result can be summarized as follows.
Let $\{{\cal S}_i(p)\}$ be one-parameter families of solutions of
Einstein's  equations for spherically symmetric massless scalar field
minimally coupled to gravitational field, where $p$ is a parameter
which smoothly specifies the initial value of the scalar and
gravitational fields and $i$ is an index which specifies one of the
several one-parameter families. For each family $S_i(p)$ there is a
critical value $p^*$ of the parameter $p$ such that solutions
$S_i(p>p^*)$ contain a black hole while $S_i(p<p^*)$ do not, 
which are referred to as supercritical and subcritical, respectively.
He found that near-\crit\ solutions satisfy the following:
(1) the strong field region is universal in the sense that it  
approaches the identical spacetime for all 
families (\ie for all  $i$),
(2) the strong field region has a discrete self similarity, and 
(3) for \superc\ solutions the \bh\ mass is proportional to 
a certain power of the deviation of 
the parameter $p$ from the \crit\ value $p^*$ as 
\begin{equation}
\MB\propto (p-p^*)^\beta,     
\end{equation}
and the \crit\ exponent $\beta$ is universal, \ie independent of $i$. 
The numerical value of the $\beta$ was given as $\beta\simeq0.37$, and
was speculated to be $1/e$. 
Abrahams and Evans \cite{AbEv93} found the similar phenomena in 
axisymmetric collapse of gravitational wave with almost the same value
of the \crit\ exponent $\beta\simeq0.38$. Evans and Coleman
\cite{EvCo94} also found the similar phenomena with $\beta\simeq0.36$ 
in spherically symmetric collapse of radiation fluid,
in which case the self similarity is not 
discrete but local one. The difference seems to come from that in the
first two cases the matter propagates as waves while in the last case
it does not. 
There might be some typical wavelength in the first two cases.
It should be noted that these calculations may suggest that the
meaning of the universality in (3) can be extended to independence of
the critical exponent from details of systems, though in the first
suggestion it meant its independence of initial data. 

The phenomena described above are very striking, but there are almost
no theoretical account for the essense of them, especially for the 
value of the critical exponent $\beta$ and 
the mechanism to accquire the self similarity and the universality.
 Oshiro, Nakamura and Tomimatsu \cite{ONT94}, 
Husain, Martinez, and Nunez \cite{HMN94},
and Brady\cite{Bra94} 
studied the critical behavior of scalar field collapse 
using the exact self-similar solution.
They gave  the the critical exponent
$\beta<\hspace{-12pt}{}_\sim\;0.5$. 
Their argument on $\beta$ was not very quantitative, 
which was mostly due to the difficulty of estimating the form of 
the metric in the asymptotic region.

If the critical behavior has very universal feature
we may expect the similar  behavior in seemingly very different 
models, which may enable us to investigate it in a simple model.
On the other hand, the model should have some essential 
properties which the above models have in common. 
First, the spacetime should be asymptotically flat.
Second, there must be energy transportation to the null infinity.
This is because otherwise the spacetime necessarily ends up with a
black hole with the mass of the initially prepared matter whenever a
black hole forms (apart from the possibility that a star forms outside
a black hole of small mass) so that there would be no criticality, 
which is the case of the Lemaitre--Tolman--Bondi spacetimes.
Third, the matter content was ``massless'' in all of the above
numerical studies. It is preferable that our model should also have
the similar property.

A simple model possessing these properties is the collapse of a thin
shell coupled with outgoing null flux.  
We will find such a model which exhibits the critical 
behavior in \bh\ formation.
We treat it analytically, which may help find the mechanism of the 
critical behavior.
We shall find the critical exponent 
and its dependence on the details of the systems.

Sec. \ref{general} is a brief review of the treatment of a thin shell
in general relativity. In sec. \ref{our model} we present the model of
thin shell coupled with null flux. We find its critical behavior in sec. 
\ref{critical}. 
In sec. \ref{approx}, approximate solutions of the dynamics of the thin
shell is derived and the value of the critical exponent is calculated by
using them.  
The exactness of the value is rigorously demonstrated in Sec.
\ref{exact}. 
Sec. \ref{conc} is devoted to summary, and Sec. \ref{disc} to
discussion. 

\section{General treatment of dynamics of a thin shell}
\label{general}
We briefly summarize the prescription of dynamics of an infinitely
thin shell using metric junction which was first formulated by
Israel\cite{Isr66}. 
Let $S$ be a compact 2-surface which represents the thin shell. 
The shell's trajectory is given by two embedding of $S\times\R$ into
the {\em exterior}\/ and the {\em interior}\/ spacetimes $V_+$ and
$V_-$, which are supposed to be the solution of Einstein's equation. 
Let $n^a$ denote the outward-pointing unit normal vector to the
trajactory and $h_{ab}$ denote the induced metric on the trajectory.  
The nontrivial components of Eintein's equation on the shell are 
\begin{eqnarray}
        G_{ab}n^an^b&=& 8\pi T_{ab}n^an^b,
        \label{eq:i} \\
        G_{ab}n^ah^b{}_c&=& 8\pi T_{ab}n^ah^b{}_c,
        \label{eq:ii}
\end{eqnarray}
which are equivalent to 
\begin{eqnarray}
        {}^{(3)}R+K_{ab}K^{ab}-K^2 &=& -16\pi T_{ab}n^an^b,
                \label{eq:iii}\\
        D_bK^b{}_c- D_cK &=& 8\pi T_{ab}n^ah^b{}_c,
                \label{eq:iv}
\end{eqnarray}
where ${}^{(3)}R$ denotes the 3-curvature of the trajectory and $D$
denotes the covariant derivative with respect to the metric $h_{ab}$. 
The difference of the above equations 
between $V_+$ and $V_-$ is interpreted as due to the
energy-momentum of the shell:
\begin{eqnarray}
        \tK_{ab}S^{ab} &=& [T_{ab}n^an^b],
        \label{eq:v}\\
        D_bS^b{}_c &=& -[T_{ab}n^ah^b{}_c],
        \label{eq:vi}
\end{eqnarray}
where $\tK_{ab}=(1/2)(K^+_{ab}+K^-_{ab})$, $[A]=A_+-A_-$, and $S_{ab}$
is defined as $S_{ab}=(1/8\pi)[h_{ab}K-K_{ab}]$ which is interpreted
as the energy-momentum of the shell. 

We assume that the shell
has the surface energy density $\sigma$ 
and tension $\zeta$, \ie,
\begin{equation}
        S_{ab}=\sigma u_au_b-\zeta (h_{ab}+u_au_b),
        \label{eq:shell's EM}
\end{equation}
where $u^a$ is a timelike unit vector along the shell's trajectory.
We also assume that the shell is spherically symmetric.
Then the parts tangent to and normal to the vector $u^a$ of 
eq. (\ref{eq:vi}) are
\begin{eqnarray}
        u^bD_b\sigma+{2u^bD_bR\over R}(\sigma-\zeta)&=&[T_{ab}n^au^b],
        \label{eq:EC} \\
        u^bD_bu_c &=&0,
        \label{eq:particle}
\end{eqnarray}
respectively, where $R$ is the area radius of the shell.
Eq. (\ref{eq:EC}) is the equation for the energy flow. 
Eq. (\ref{eq:particle}) is the equation of motion of the fluid
particle of the shell, which indicates that each particle moves along
the geodesic on the three-dimensional  
hypersurface of the trajectory of the shell.

 From the definition of $S_{ab}$ one has
\begin{eqnarray}
        \sigma&=& -{1\over8\pi}(u^cu^d+h^{cd})[\nabla_cn_d]\nonumber\\
                  &=& -{1\over4\pi}{[n^d\nabla_d r]\over R},
                \label{eq:defS}
\end{eqnarray}
where $r$ is the area radius of the symmetric sphere and 
$[n^d\nabla_d r]$ is evaluated at $r=R$.
In terms of the shell's proper mass ${\cal M}=4\pi\sigma R^2$ it can
be written as  
\begin{equation}
-{{\cal M}\over R}=[n^d\nabla_d r].
        \label{}
\end{equation}
The equation for the tension $\zeta$ can be also derived, 
but it is not an independent equation in the model presented below.
The model also satisfies eq. (\ref{eq:v}).
Eqs. (\ref{eq:EC}) and (\ref{eq:defS}) determine the motion of the 
shell.  

\section{Collapse of a thin shell coupled with null flux}
\label{our model}
We assume that the outer side of the shell is coupled with outgoing
null flux which may transport the energy of the shell to infinity
and the inner side of the shell does not have any interaction.
The exterior spacetime is the Vaidya spacetime. 
The interior spacetime is assumed to be flat, \ie, is the Minkowski 
spacetime. 
We can write the line element  as
\begin{equation}
        ds^2=-(1-2\Phi) du^2-2dudr+r^2d\Omega^2,
\end{equation}
where $u$ is the retarded time, $r$ the area radius, and $d\Omega^2$ 
the line element on a unit sphere,
and
\begin{equation}
\Phi=\left\{
\begin{array}{ll}
 \Phi_+={M(u)/ r}  &({\rm exterior})\\
 \Phi_-=0             &({\rm interior}).
\label{eq:Phi}
\end{array}
\right.
\end{equation}
A similar situation is considered in \cite{MSY94} in the context of 
the creation of man-made universes. 
Note that $M(u)$ gives the Bondi--Sachs mass of the null infinity at
$u$. 

Let $\tau$ denote the proper time of the shell.
Let $r=R(\tau)$ represent the trajectory of the shell. 
Note that $R$ is the shell's radius. The $r$-component 
of the vector $u^a$ is written as $u^r=\dot R$, where a dot denotes
the differentiation with respect to $\tau$.
It follows from the normalization condition that 
the $u$-component is written as $u^u=(\sqrt{\dR^2+1-2\Phi}+\dR)^{-1}$. 
(Hereafter $\Phi$ denotes $\Phi_+(u,R)$.)
Then the vector $n$ is given as $n^\alpha=s
(-u^u,\sqrt{\dR^2+1-2\Phi})$,  
where $s $ is plus/minus unity if $r$ is increasing/decreasing in 
the direction of $n^a$.
Eq. (\ref{eq:defS}) yields
\begin{equation}
  \dR^2+1=f^2,
\label{eq:ee1}
\end{equation}
where
\begin{equation}
  f={1\over2}\paren{{\cal M}\over R}+
    \Phi\paren{{\cal M}\over R}^{-1}.
\label{eq:deff}
\end{equation}
The outgoing flux on the shell is
\begin{equation}
[T_{ab}n^au^b]=  
\left\{
\begin{array}{ll}
  s {\dM u^u/ 4\pi R^2}&({\rm exterior})\\
         0             &({\rm interior})
\end{array}
\right.
\end{equation}
where we used $\dM=M_{,u}u^u$.
Then eq. (\ref{eq:EC}) yields
\begin{equation}
        \dot\sigma+{2\dot R\over R}(\sigma-\zeta)=s \, {\dM\over 4\pi 
          R^2} {1\over \sqrt{\dR^2+1-2\Phi}+\dR}.
        \label{eq:ee2}
\end{equation} 

We need two more conditions to specify the system completely. 
One is the equation of state of the shell and the other is the
coupling of the shell and the spacetime. We assume that  the former is
given by  
\begin{equation}
\zeta=-\eta \sigma
\end{equation}
with $\eta$ a constant.
For simplicity we assume that the latter is given by
\begin{equation}
(\sigma R^{1-\omega})^\cdot=0,      
\end{equation}
\ie,
\begin{equation}
\sigma R^{1-\omega}=C^\omega/4\pi,
\label{eq:coupling}
\end{equation}
where $\omega$ and $C$ are constants. 
In the $\omega=-1$  case the mass $M$ of the shell conserves, 
while  $\omega=1$ is said to hold in the case of false vacuum bubbles
\cite{BGG87}. 
So our assumption is a slight generalization of them in the 
parameter space. 
 From eqs. (\ref{eq:deff}) and (\ref{eq:coupling}) one has
\begin{equation}
  f=\tR^{-\omega-1}\tM+\tR^\omega/2,
\label{eq:f}
\end{equation}
and eq. (\ref{eq:ee2}) yields
\begin{equation}
\dot{\tM}= s \, p\, \dot{\tR}\,\tR^\omega\,
(\sqrt{{\dR}^2+1-2\Phi}+{\dR}), 
\label{eq:ee2-2}
\end{equation}
where $p=2\eta+\omega+1$, $\tM=CM$,$\tR=CR$, and $\Phi$ is given in  
eq. (\ref{eq:Phi}).
Eqs. (\ref{eq:ee1}) and (\ref{eq:ee2-2}) are the basic equations in
the subsequent discussions.

\section{Model with critical behavior}
\label{critical}
Let us look at the qualitative behavior of the shell and search models
which exhibits a critical behavior in the \bh\ formation.
First we assume that there is some mechanism to prevent negative flux
to emit from the shell and the shell will be static after then.
It follows from eqs. (\ref{eq:ee1}), (\ref{eq:ee2-2}) and ${d\tM/ 
d\tR}={\dot{\tM}/ \dot{\tR}}$ that the orbit of the shell outside the
event horizon ($s=1$) is given by a first order ordinary differential
equation 
\begin{equation}
\label{eq:ode}
{d\tM\over d\tR}=p \tR^\omega\, (f-\tR^\omega+
    {\dot{\tR}\over|\dot{\tR}|}\sqrt{f^2-1}),
        \label{eq:orbit}
\end{equation}
where $f$ is given by eq. (\ref{eq:f}).

One immediately finds in the above equation 
that the value of $C$, which is one of the 
parameters specifying the initial condition, is irrelevant to the  
qualitative behavior of the shell, \ie, the motion of the shell is
scale invariant. This is another feature which our model has in common
with the case of spherically symmetric scalar field collapse.
The set $T$ of turning points (boundary of the forbidden region)
in the $\tM$-$\tR$ diagram is given by the points at which
$\dot{\tR}=0$. 
 From eq. (\ref{eq:f}) this implies $f=1$ so that the mass of the shell
at a point on $T$ is expressed as
\begin{equation}
\tMT=\tR^{\omega+1}\paren{1-{\tR^\omega\over2}}.
\label{eq:tp}
\end{equation}
The line $H$ of $\tM=\tR/2$ 
corresponds to the apparent horizon where the
expansion of the outgoing null ray vanishes. Further, it coincides
with the event horizon for the outgoing Vaidya spacetime. One can
easily check that $H$ is tangent to $T$ (if $\omega\neq0$).
The shape of $T$ depends on the sign of $\omega$,
which is illustrated in fig. 1.
If $\omega>0$ then $T$ passes the origin and the mass of the shell is
bounded from below to be positive.
If $\omega<0$ the mass is not bounded from below. This implies that
the mass of the shell can be negative at $\tR=0$, which may suggest
that a naked singularity occurs. We shall not concern the critical
point between black hole and naked singularity here and assume that
$\omega$ is positive. 
The initial condition $(\tRI, \tMI)$ can be given in the region
between $H$ and $T$ with $0\le\tR<1$, which we will call Region I, 
if the shell is not contained in a black hole from the beginning.

One would like to consider the situation in which the shell's energy
is carried away to infinity during the collapse. 
We therefore require that the mass decreases when the shell is
collapsing. This is given by the condition
$
p=2\eta+\omega+1>0,
  \label{eq:sign of mdot}
$
because in Region I we always have $\sqrt{f^2-1}\ge f-\tR^\omega\ge0$.
Then a collapsing shell has only three possibilities:
it will hit (1) $H\, (\tR>0)$, (2) $T\, (\tR>0)$, or (3) the origin. 
The case (1) implies that there is a formation of a \bh; the solution
is supercritical. 
Since the local gravitational mass
is constant along the outgoing null line, the mass $M_B$ of the black
hole, which is given by the limit of the Bondi mass to the future
timelike infinity $i^+$, 
is given by the mass of the shell crossing the event horizon $H$.
The case (2) implies that the shell reaches the turning point and
expands. But eq. (\ref{eq:orbit}) implies that negative flux emerges
when it expands so that the shell will stay at the turning point. 
The solution is subcritical. 
If both (1) and (2) are possible there may be a
solution of the case (3), which should give the critical point.

Whether each case really occurs can be examined by using the gradient 
${d\tM/d\tR}$ given by (\ref{eq:orbit}).
On $H$ one has ${d\tM/d\tR}=0$, which implies that \bh\
solutions always exist. The gradient is always positive in Region I
and negative outside Region I. Therefore, the shell loses/gains mass
before/after it cross the event horizon.
In Region I the gradient monotonically
increases when the point is going down to $T$ along the constant $\tR$
line, because the partial derivative of the right hand side of
(\ref{eq:orbit}) by $\tM$ is always negative in Region I. 
For each point of $H$, one can check whether there is a solution
with decreasing $\tR$ which hits it 
by comparing the gradient of the solution and that of the line $T$.
The difference is given by
\begin{eqnarray}
  \paren{d\tM\over d\tR}_{{\rm on \ }T}-
  {d\tMT\over d\tR} &=&-2\tR^\omega
  \paren{\eta-\paren{\eta+{1\over4}}\tR^\omega} 
\end{eqnarray}
The point at which this is positive (negative) is the turning point
 from contraction to expansion (expansion to contraction). 
One finds that it changes sign once from positive to negative 
at $\tR=\tRS =(\eta/(\eta+1/4))^{1/\omega}$ in
$0<\tR<1$ if $\eta>0$, and it is always negative otherwise.
This implies that solutions of the above case (2) certainly exists
if and only if $\eta$ is positive.  Combining these conditions,
we find that the critical point exists if and only if 
the both of $\eta$ and $\omega$ are positive.
A few remarks are now in order. 
First, the condition $\eta>0$ implies that the intrinsic force of the 
shell is not tension but pressure. This is a reasonable condition
for the collapsing shell to expand. 
Second, all solutions with decreasing $\tR$ has its ``starting point''
on $T$ where $\tR\ge\tRS$. The ``terminal point'' is either on $H$ or
on $T$ with $\tR<\tRS$.
Third, the point $\tR=\tRS$ on $T$ is the solution of the static
shell. 

The critical solution $\tM=\tMC$ (\ref{eq:orbit}) is the one which
passes the origin. It is easily found that
$\tMC\propto\tR^{1+\omega}+o(\tR^{1+\omega})$, in particular, the
critical solution is tangent to $T$ at the origin.
As already said, the \bh\ mass $\tMB$ is given by the value of $\tM$ 
at the intersecting point of the solution and  $\tM=\tR/2$.

The Penrose diagrams of the future half of each case are shown in fig.
2. In the critical solution the asymptotic behavior
$\tMC\propto\tR^{1+\omega}$ implies that the invariant
$R_{abcd}R^{abcd}=48M^2/r^6$ diverges when the shell shrink to $r=0$
if $0<\omega<2$, while it is finite if $\omega\ge2$. 
The former case can be interpreted that the shell collapsed to a naked
singularity in the center, while the latter that the shell evaporated.
In both cases the curvature is always continuous in $r>0$.

\section{Approximate solutions}
\label{approx}
In this section we consider the motion of 
the shell in some datail and obtain approximate solutions by an 
intuitive argument.
We shall calculate the critical exponent by using them.
We shall give a more rigorous argument in the next section.

We change variables for easier manipulation of the differential
equation (\ref{eq:orbit}) by
\begin{eqnarray}
  x&=&\tR^\omega \\ y&=&2\tM\tR^{-1}.
  \label{eq:change}
\end{eqnarray}
Eq. (\ref{eq:orbit}) for $\dot{\tR}<0$ is given by
\begin{equation}
 {dy\over dx}= {1\over\omega}\paren{(p-1){y\over x}-px-
    p\paren{\paren{{y\over x}+x}^2-4}^{1\over2}}.
\label{eq:dy/dx}
\end{equation}
The behavior of the solutions is illustrated in fig. 3.
The set $T$ of turning points is given by the parabolic curve
$y=x(2-x)$. The critical solution $y=y_C$ in $y$-$x$ diagram hits the
origin, because $\tMC\propto\tR^{1+\omega}+o(\tR^{1+\omega})$ implies
that $y_C/x\rightarrow0$ as $x\rightarrow0$.
On the other hand, the \superc\ solutions diverge at $\tR=0$.
The event horizon $H$ is given by the straight line $y=1$.
The mass of the black hole $\tM_B$ is related to the value $x_B$ of
$x$ at the point intersecting $H$ as $\MB\propto\xB^{1/\omega}$.
A convenient choice for initial conditions of $(\tR,\tM)=(\tRI,\tMI)$
is those with fixed $\tR$, which correspond to
initial conditions of $(x,y)=(\xI,\yI)$ with fixed $\xI$.
By definitions of $x$ and $y$ one has 
a relation $\DtMI\propto \DyI$ for the initial values,
where $\Delta$ denotes the difference from the critical solution.
Then a power-law relation $\xB\propto\DyI^{\beta'}$ implies 
$\tMB\propto\DtMI^\beta$, where $\beta=\beta'/\omega$.
We look for the value $\beta'$ in the following.

Let us devide Region I into two, $x\ge\xcut$ and $x\le\xcut$,
with $\xcut$ being a small positive constant.
We will call the first region the {\em linear (deviation) region}.
In the linear region,
a small deviation from the critical solution is always linear, \ie
\begin{equation}
\Delta\ycut\propto\DyI. 
\label{eq:cut vs I}
\end{equation}
It is because in  $x>\xcut$ the rate of change of the gradient,
 $(\partial/\partial y)(dy/dx)$, is bounded so that the deviation
 vector is always finite. 
It follows that nonlinearity of the deviation of the Bondi mass from 
the critical solution emerges only near $x=0$.

The region $x\le\xcut$ will be called the {\em scaling region}. If we  
ignore $x$ compared to unity, eq. (\ref{eq:dy/dx}) reads 
\begin{equation}
 {dy\over dx}= {1\over\omega}\paren{(p-1){y\over x}-
    p\paren{\paren{{y\over x}}^2-4}^{1\over2}}.
\label{eq:dy/dx2}
\end{equation}
Note that $y/x$ is not necessarily large compared to unity for
near-critical solutions.
One finds a scaling law in the sense that the transformation
$(x,y)\mapsto(ax,ay)$ does not change the differential equation
(\ref{eq:dy/dx2}),  which says that if
$y=y_1(x)$ is a solution, $y=a^{-1}y_1(a x)$ gives general solutions.
In fact, one integrates eq. (\ref{eq:dy/dx2}) to find
\begin{equation}
k=x\,(\sqrt{f^2-1}+f)^{\omega p \over p^2-q^2}
\left|p \sqrt{f^2-1} -q f\right|^{\omega q \over p^2-q^2},
\label{eq:f0}
\end{equation}
where $q=2\eta$, $k$ is the constant of integration, and 
$2f=y/x$ in the present approximation. 
The critical solution is the one passing the origin hence 
the one with $f=y/2x$ being finite when $x\rightarrow0$.
It is the $k=0$ solution, namely
\begin{equation}
  y_C={2px\over\sqrt{p^2-q^2}},
\end{equation}
Positive $k$ corresponds to all of the other solutions.
The supercritical/subcritical solutions are the ones with 
the term between the vertical lines in (\ref{eq:f0})
positive/negative, because $f$ should be greater/less than the
critical solution $f_C$. For small $k$, by setting $y=1$ one finds 
\begin{equation}
        k\sim\xB^{1-{\omega\over p-q}}.
        \label{eq:0 vs B}
\end{equation}
Setting $x=\xcut$ one has
\begin{equation}
        k\sim(\Delta\ycut)^{\omega q\over p^2-q^2},
        \label{eq:0 vs I}
\end{equation}
where one sees that the power is determined by the power of the
term between the vertical lines in (\ref{eq:f0}).
 From eqs. (\ref{eq:cut vs I}), (\ref{eq:0 vs B}), (\ref{eq:0 vs I})
and $\beta=\beta'/\omega$ one obtains the relation
$\tMB\propto\tMI^\beta$ with the critical exponent 
\begin{equation}
        \beta={q\over p+q}={2\eta\over 4\eta+\omega+1}.
        \label{eq:beta}
\end{equation}

\section{Exactness of the critical exponent}
\label{exact}
In the previous section we found the value of the critical exponent in
a very simple approximation. 
However, it can be shown that the value actually is {\em exact} by a 
more rigorous treatment of eq. (\ref{eq:ode}). 
The idea is that the cutoff scale $\xcut$ now is dependent on the
initial deviation of $\yI$ such that $\xcut$ tends to zero when
$\Delta\yI\rightarrow0$. Several points with different velocities of
convergence divide the solution curve into a corresponding number of
segments.  We shall find a nonlinear 
relation of the deviations on each interval which eventually gives 
the value of $\beta$.

We rewrite eq. (\ref{eq:dy/dx}) in terms of $f=(1/2)(y/x+x)$ 
instead of $y$ to have
\begin{equation}
{df\over dx}={1\over\omega x}(qf-p\sqrt{f^2-1}-(q+{1\over2})x).
\label{eq:df/dx}
\end{equation}
%The inequality $f>(1/2)(1/x+x)$ implies the formation of a black
%hole. 
Let us briefly show the existence and uniqueness of the critical
solution, a solution which has a finite limit at $x\rightarrow0$. 
We consider solutions of (\ref{eq:df/dx}) in 
$U=(0,\delta)\times
(f_{C0}-(2q+1)\delta/\omega, f_{C0}+(2q+1)\delta/\omega)$ 
where $f_{C0}=p/\sqrt{p^2-q^2}$ and $\delta$ is a positive constant 
(fig. 4).
It can be easily shown that there exists $\delta$ such that 
$df/dx\le\mbox{constant}<0$ on line
$l_1:f=f_{C0}\in U$ and 
$df/dx-(-(2q+1)x/\omega )\ge\mbox{constant}>0$ on 
$l_2:f=f_{C0}-(2q+1)x/\omega\in U$.
A solution cannot pass both of $l_1$ and $l_2$ in $U$.
Moreover, sets of points of solutions passing $l_1$ and of those of
solutions passing $l_2$ form open sets in $U$.
This implies that there must be a point which does not cross neither
$l_1$ and $l_2$ in $U\cap\{(x,f); f_{C0}-(2q+1)x/\omega<f<f_{C0}\}$.
So there must exist a solution $f=f_C$ which satisfies
\begin{equation}
f_{C0}-{(2q+1)x\over\omega}\le f_C \le f_{C0}
\label{eq:<fC0}
\end{equation}
in $0<x<\delta$ and has a limit $f_C\rightarrow f_{C0}$ as
$x\rightarrow0$. 

Let us define $\Delta$ by $\Delta=f- f_C$. 
We consider solutions with nonnegative $\Delta$.
The equation for $\Delta$ is given by
\begin{equation}
{d\ln\Delta\over d\ln x}
=-{1\over\omega}(p F(f, f_C)-q),
\label{eq:**}
\end{equation}
where
\begin{equation}
F(a,b)=F(b,a)={a+b\over \sqrt{a^2-1}+\sqrt{b^2-1}}
\end{equation}
for $a,b>1$.

We show some inequalities derived from eq. (\ref{eq:**}).
$F(a,b)$ is a decreasing function of $a$ and $b$ so that 
\begin{equation}
F(a, b)\ge F(c, c)={c\over\sqrt{ c^2-1}}\ge1
\end{equation}
holds for $c\ge a,b$.
Since $f\ge f_C>1$, if $f\le c$ on interval $[\xmin,\xmax]$ then
\begin{equation}
{d\ln\Delta\over d\ln x}\le
-{1\over\omega}\left({p c\over\sqrt{c^2-1}} -q \right)
\le -{p-q\over \omega}
\end{equation}
holds on the interval. Its integration on interval $[\xmin,\xmax]$
gives 
\begin{eqnarray}
{\Delta (\xmin)\over\Delta (\xmax)}
&\ge& \left({\xmax\over \xmin}\right)^
      {{1\over\omega}\left({p c\over\sqrt{c^2-1}}-q\right)}
\label{eq:1a}\\
&\ge& \left({\xmax\over \xmin}\right)^{p-q\over\omega}.
\label{eq:1b}
\end{eqnarray}
In particular, $\Delta$ increases as $x$ decreases and diverges as 
$x\rightarrow0$. This implies that the critical solution is unique.

By this and a similar argument for negative $\Delta$, one finds that 
all solutions $f>f_C$ go to $\infty$ as $x\rightarrow0$, while
ones $f<f_C$ to $-\infty$ (though, in fact, $f<1$ is forbidden).
This tells that the point $(0,f_{C0})$ is a {\em separatrix}.
One expects that the deviation from the critical solution becomes
nonlinear near the separatrix.

If $d\le f_C<f\le c$ on the interval $[\xmin,\xmax]$ 
with $c-d$ being sufficiently small then
\begin{equation}
F(f,f_C)\le {d\over \sqrt{d^2-1}}
\le{c\over \sqrt{c^2-1}}+{c-d\over (d^2-1)^{3/2}}
\end{equation}
holds, where the second inequality is because $c/\sqrt{c^2-1}$ is a
convex function of $c$. This gives
\begin{equation}
{d\ln\Delta\over d\ln x}\ge -{1\over\omega} 
\left({pc\over\sqrt{c^2-1}}-q+{p(c-d)\over (d^2-1)^{3/2}}\right),
\end{equation}
and its integration yields
\begin{equation}
{\Delta (\xmin)\over \Delta (\xmax)}\le
\left({\xmax\over\xmin}\right)^
{{1\over\omega}
\left({pc\over\sqrt{c^2-1}}-q+{p(c-d)\over (d^2-1)^{3/2}}\right)}.
\label{eq:2}
\end{equation}
If $f_C\le d<c\le f$ on $[\xmin,\xmax]$ with
sufficiently large $c$ and fixed $d$ it is easily shown that 
\begin{equation}
F(f,f_C)\le 1+{d+1\over c}
\end{equation}
holds on $[\xmin,\xmax]$, so that one has
\begin{equation}
{\Delta (\xmin)\over \Delta (\xmax)}\le
\left({\xmax\over\xmin}\right)^
{{1\over\omega}\left(p-q+{p(d+1)\over c}\right)}.
\label{eq:3}
\end{equation}
 
%%%%%12
For notational simplicity we adopt the convention that subscript  
$i(=1,2,3,...)$ denote the value at $x=x_i$ and subscript 0 denotes
the value at $x=0$.
Let $x_1=\xI$ and $\Delta_1=\epsilon$ (fig. 5).
Let us define $x_2=|\ln\epsilon|^{-1}$.
 From eq. (\ref{eq:1b}) one has
\begin{equation}
{\Delta_2\over\Delta_1}
\ge
\left({ x_1\over|\ln\epsilon|^{-1}}\right)^{p-q\over\omega}.
\label{eq:12}
\end{equation}
 From eq. (\ref{eq:**}) one has
\begin{eqnarray}
{d\ln\Delta\over d\ln x}
&= &
-{1\over\omega} 
\left({p(2f_C+\Delta)\over \sqrt{f^2-1}+\sqrt{f_C{}^2-1}}-q\right) 
\nonumber\\
&\ge &
-{1\over\omega} 
\left({pf_{C1}\over\sqrt{ f_{C1}^2-1}}-q+
{p\Delta\over2\sqrt{ f_{C1}^2-1}}\right),
\end{eqnarray}
\ie,
\begin{equation}
{1\over P\Delta(1+Q\Delta)}{d\Delta\over d\ln x}\ge -1,
\end{equation}
where
$P$ and $Q$ are positive constants. Integration on $[x_2,x_1]$ gives 
\begin{equation}
{1+{1\over Q\Delta_2}\over1+{1\over Q\Delta_1}}\ge
\left(x_2\over x_1\right)^P,
\end{equation}
hence
\begin{eqnarray}
{\Delta_2\over\Delta_1}
&\le&\left(\left({x_2\over x_1}\right)^P-
             Q\Delta_1\left(1-\left({x_2\over x_1}\right)^P
             \right)\right)^{-1}\nonumber\\
&=&\left(\left({|\ln\epsilon|^{-1}\over x_1}\right)^P-
             Q\epsilon\left(1-\left({|\ln\epsilon|^{-1}\over 
             x_1}\right)^P\right)\right)^{-1}\nonumber\\             
&\le&\left(\left({|\ln\epsilon|^{-1}\over x_1}\right)^P-
             {1\over2}\left({|\ln\epsilon|^{-1}\over 
             x_1}\right)^P\right)^{-1}\nonumber\\   
&=&2\left({x_1\over|\ln\epsilon|^{-1}}\right)^P,
\label{eq:11}
\end{eqnarray}
where the inequality between the second and the third lines holds for
sufficiently small $\epsilon$.
Combining inequalities (\ref{eq:12}) and (\ref{eq:11}) one has 
\begin{equation}
\Delta_2\sim\Delta_1\sim\epsilon.
\end{equation}
Here $A\sim B$ means that $A$ and $B$ are the same in power of
$\epsilon$ up to logarithm. More precisely, $A\sim B$ holds if 
there exist positive constants, \ie numbers independent of $\epsilon$,
$P,P'$, and constants $Q,Q'$ such that 
\begin{equation}
P|\ln\epsilon|^Q\le{A\over B}\le P'|\ln\epsilon|^{Q'}
\end{equation}
holds.

%%%%%23
Let us define $x_3$ such that $\Delta_3=|\ln\epsilon|^{-1}$.
 From $\Delta_2\sim\epsilon$ one has $x_3\le x_2$ for sufficiently 
small $\epsilon$. 
Inequality (\ref{eq:<fC0}) and the definition of $x_3$ implies that 
\begin{eqnarray}
f_{C0}-{2q+1\over\omega}|\ln\epsilon|^{-1}
\le & &f_C\le f_{C0},\nonumber\\
& &f_C\le f\le f_{C0}+|\ln\epsilon|^{-1}.
\label{eq:f on 23}
\end{eqnarray}
It follows from eqs. (\ref{eq:f on 23}) and (\ref{eq:1a}) that,
if $\epsilon$ is sufficiently small,
\begin{equation}
{x_3\over x_2}\ge 
\left({\Delta_2\over \Delta_3}\right)^
{{\omega q\over p^2-q^2}+P|\ln\epsilon|^{-1}}
\label{eq:620}
\end{equation}
holds for some constant $P$.
 From inequalities (\ref{eq:f on 23}) and (\ref{eq:2}) 
\begin{equation}
{x_3 \over x_2}\le
\left({\Delta_2\over \Delta_3}\right)^
{{\omega q\over p^2-q^2}+P'|\ln\epsilon|^{-1}}
\label{eq:621}
\end{equation}
holds for some constant $P'$.
It follows from these inequalities 
and the orders in $\epsilon$ of 
$x_2$, $\Delta_2$ and $\Delta_3$ 
that
\begin{equation}
x_3\sim \epsilon^{\omega q\over p^2-q^2},
\label{eq:x3}
\end{equation}
because the term $P'|\ln\epsilon|^{-1}$ in the exponent causes only 
logarithmic corrections in (\ref{eq:621}). 

%%%%%34
Let us define $x_4$ such that $\Delta_4=|\ln\epsilon|$.
Monotonicity of $\Delta$ implies 
\begin{equation}
  x_4\le x_3.
\end{equation}
Eq. (\ref{eq:1b}) implies 
\begin{equation}
{x_4\over x_3}\ge
\left({\Delta_3\over\Delta_4}\right)^{\omega\over p-q}.
\end{equation}
 From these inequalities one has
\begin{equation}
x_4\sim x_3\sim \epsilon^{\omega q\over p^2-q^2}.
\end{equation}

%%%%%45
Let us define $x_5=x_B$. The definition of $x_B$ implies
\begin{equation}
\Delta_5={1\over2}\left({1\over x_5}+x_5\right)-f_{C5}.
\end{equation}
It can be easily checked that
\begin{equation}
f_C\le f_{C0}<\Delta_4=|\ln\epsilon|<f.
\label{eq:45ineq}
\end{equation}
 From eqs. (\ref{eq:45ineq}) and (\ref{eq:1b}) one has
\begin{equation}
{\Delta_5\over\Delta_4}\ge
\left({x_4\over x_5}\right)^{p-q\over\omega}
\end{equation}
This implies 
\begin{equation}
x_5\ge P x_4{}^{p-q}\Delta_4{}^\omega
\end{equation}
for some positive constant $P$, 
because $\Delta_5\le{1/ 2x_5}$ for sufficiently small $\epsilon$ 
and $p-q=\omega+1$.
One has from (\ref{eq:3}) that 
\begin{equation}
{\Delta_5\over\Delta_4}\le
\left({x_4\over x_5}\right)^
{{1\over\omega}\left({p-q+{P'\over |\ln\epsilon|}}\right)}
\end{equation}
for a constant $P'$.
These imply that
\begin{equation}
x_5\sim x_4{}^{p-q}\sim \epsilon^{\omega q\over p+q}.
\end{equation}
As a result, we have
\begin{equation}
x_B\sim\Delta_1{}^{\omega q\over p+q},
\end{equation}
which proves that the critical exponent $\beta$ is equal to 
${q/(p+q)}$.

\section{Summary}
\label{conc}
We constructed an analytic model which exibits a critical
behavior of black hole formation, namely, collapse of a thin shell
coupled with outgoing null fluid. 
In that model the dynamics of the shell is described by an ordinary 
differential equation. We investigated the dynamics of the shell and
the evolution of the Bondi--Sachs mass.
It was seen that 
the deviation of the Bondi--Sachs mass from that of 
the critical solution becomes 
nonlinear only near the $\tM=0$ and near the event horizon.
With suitable choice of dymamical variables $f$ and $x$, this is,
as expected in critical phenomena, 
understood as follows:
the point corresponding to the final state 
of the critical solution is represented as a separatrix,
and the nonlinear relation $\Delta M_B\propto\Delta M_I{}^\beta$ for
the supercritical solutions occurs near that point. 
We calculated the value of the critical exponent $\beta$ analytically
to find $\beta={2\eta/(4\eta+\omega+1)}$, where $\eta$ is a parameter
in the equation of state of the thin shell and $\omega$ specifies the 
strength of the coupling of the thin shell and the outgoing flux.

\section{Discussion}
\label{disc}
It is remarkable that in our model the critical exponent $\beta$ is 
dependent on the constants $\eta$ and $\omega$ specifying the system. 
The exponent $\beta$ can take the values in the range $0<\beta<1/2$ 
for $\eta>0$ and $\omega>0$. 
%We can easily adjust $\beta$ to be $\sim0.38$, but there seems to be
%no strong physical reason to prefer such values for $\eta$ and
%$\omega$. 
This suggests that the critical exponent is not so universal in the
extended sense in Sec. \ref{intro}.
There may be many different {\em universality classes}\/, and the
critical exponent $\beta$ may serve as an index specifying them.

There will be a question why the numerically calculated cases seem to
fall into the same universality class with $\beta\sim0.38$,
even if we admit the existence of many universal classes.
One may think that the massless property of the matter 
is crucial.
In our case, even if we require that the shell consists of radiation
fluid having a trace free equation of state, $\eta=1/2$, 
then $\beta$ is equal to $1/(3+\omega)$ and cannot exceed $1/3$.
This may suggest that the massless property is not so important.
Another possibility is that self-similarity is crucial. 
A support for this possibility is that the critical solution in our
model does not have spacetime self-similarity.  
If the near-critical solutions necessarily fall into a self-similar
spacetime in the systems with {\em smoothly}\/ distributed matter,
then types of self-similar spacetimes may determine the universality
classes. 

The universality in the initial value is not verified in our model
because the solutions are parametrized by a single parameter $M_I$ 
(or $R_I$), apart from an irrelevant parameter $C$. 
In our model, as said above, 
the nonlinearity of the deviation of the mass from the \crit\ solution
emerges only near the event horizon.
If the situations are similar in the numerically investigated collapses,
the universality in initial data can be naturally understood. Namely, 
almost all modes or fluctuations in initial data damp away 
so that the spacetime fall into a unique one 
by the time that the event horizon forms,
and the critical exponent depends only on that final portion
of the spacetime. 
This may also explain the universality in the extended sense
if the spacetimes fall into the same one in the last stage of the
collapse in several different systems.

To understand the universality it may be instructive to parametrize
the solution space differently, for example, $p_1=C\tMI$ and
$p_2=C^{-1}\tMI$. 
Then the critical exponent is, of course, the same for the both
parameters.  
In this point of view, one can say that the universality on
initial data is verified in our model
and that the choice of the parameters $\tMI$ and $C$
was good one in the sense that it separates the {\em relevant}\/ and
the {\em irrelevant mode}\/ to the criticality. 
In more realistic collapses, where there are many degrees of freedom,
it is expected that there is one (or a few) relevant mode
corresponding to $\tMI$ and are lots of irrelevant modes corresponding
to $C$. 
  
We remark that our model has a peculiarity that if we do not 
assume some mechanism to protect the 
energy condition and allow negative flux, the subcritical solutions
will have a reexpansion of the shell and a formation of a 
black holes of finite mass \cite{MSY94}. Though the above assumption  
seems  somewhat artificial,
we believe that in more natural models the mechanism of the emergence
of nonlinearity is similar to that of the model presented here.

Some authors have argued black hole evaporation 
by imitating the spacetimes with the Vaidya spacetime or one with
classical scalar field (e.g., \cite{His81,OKNT93}).
If these point of view have something in common with the real
machanism, some critical phenomena may be found in black hole
evaporation and would be the key to understand such spacetimes.

\section*{Acknowledgments}
We are greatly indebted to T. Hara in the
exactness proof of the critical exponent.
We are grateful to A. Hosoya for useful discussions and careful
reading of the manuscript. 
We thank S. Adachi, S. Higuchi, S. Kobayashi, H. Onozawa, and M. Siino
for fruitful conversations. 
T. K. acknowledges the Fellowship of the Japan Society for the
Promotion of Science.
This work is supported in part by the Grant-in-Aid for Scientific
Reserch Fund of the Ministry of Education of Japan (T. K.).

\newpage
\section*{Figure captions}
Fig. 1. The evolutions of the Bondi---Sachs mass and the radius of the
shell are illustrated. The line $H$ corresponds to the apparent
horizon and the line $T$ is the set of states with $\dot R=0$. 
Region I is between $H$ and $T$, where we can set initial data.
The thick line $C$ with an arrow denotes the critical solution which
hits the origin.  The thin line with an arrow 
above $C$ is a supercritical solution which cross $H$.
The value $M_B$ of the mass function is the mass of the formed black
hole. The thin line with an arrow below $C$ is a subcritical solution.

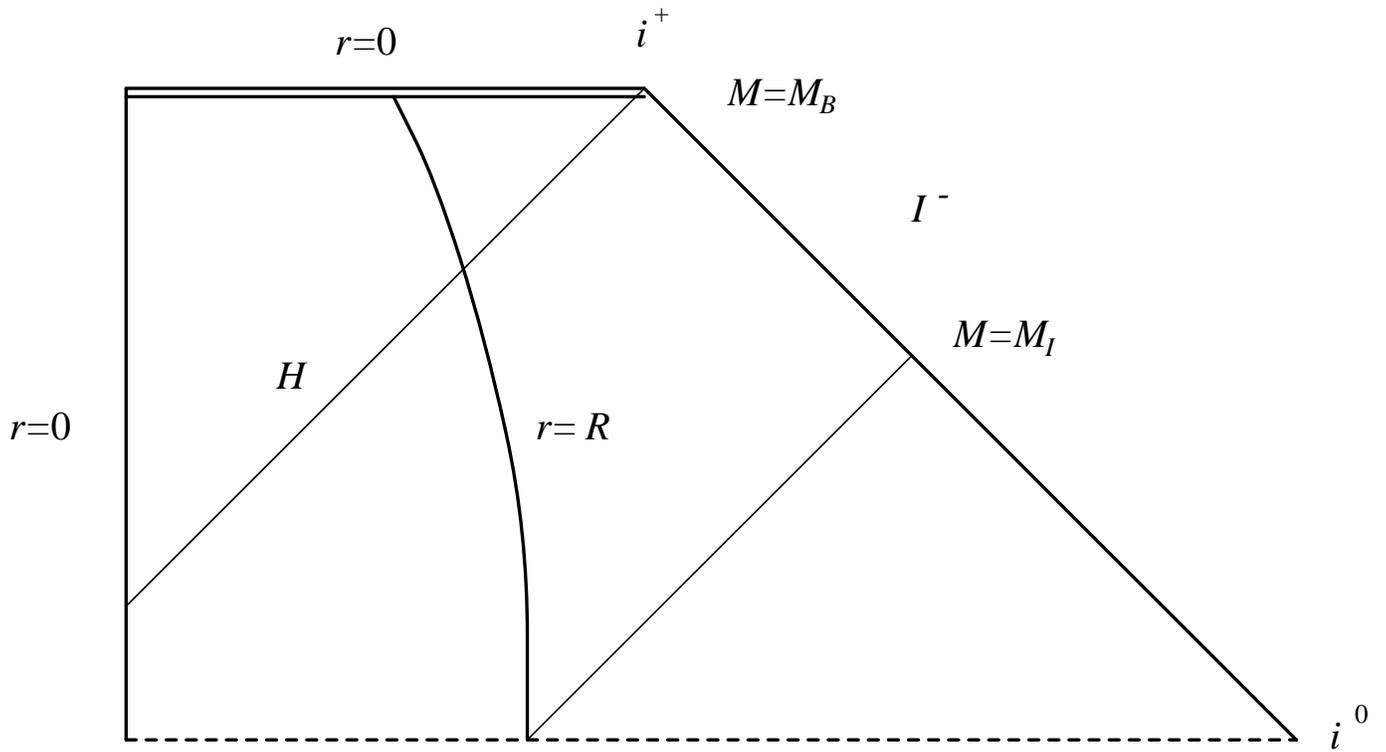
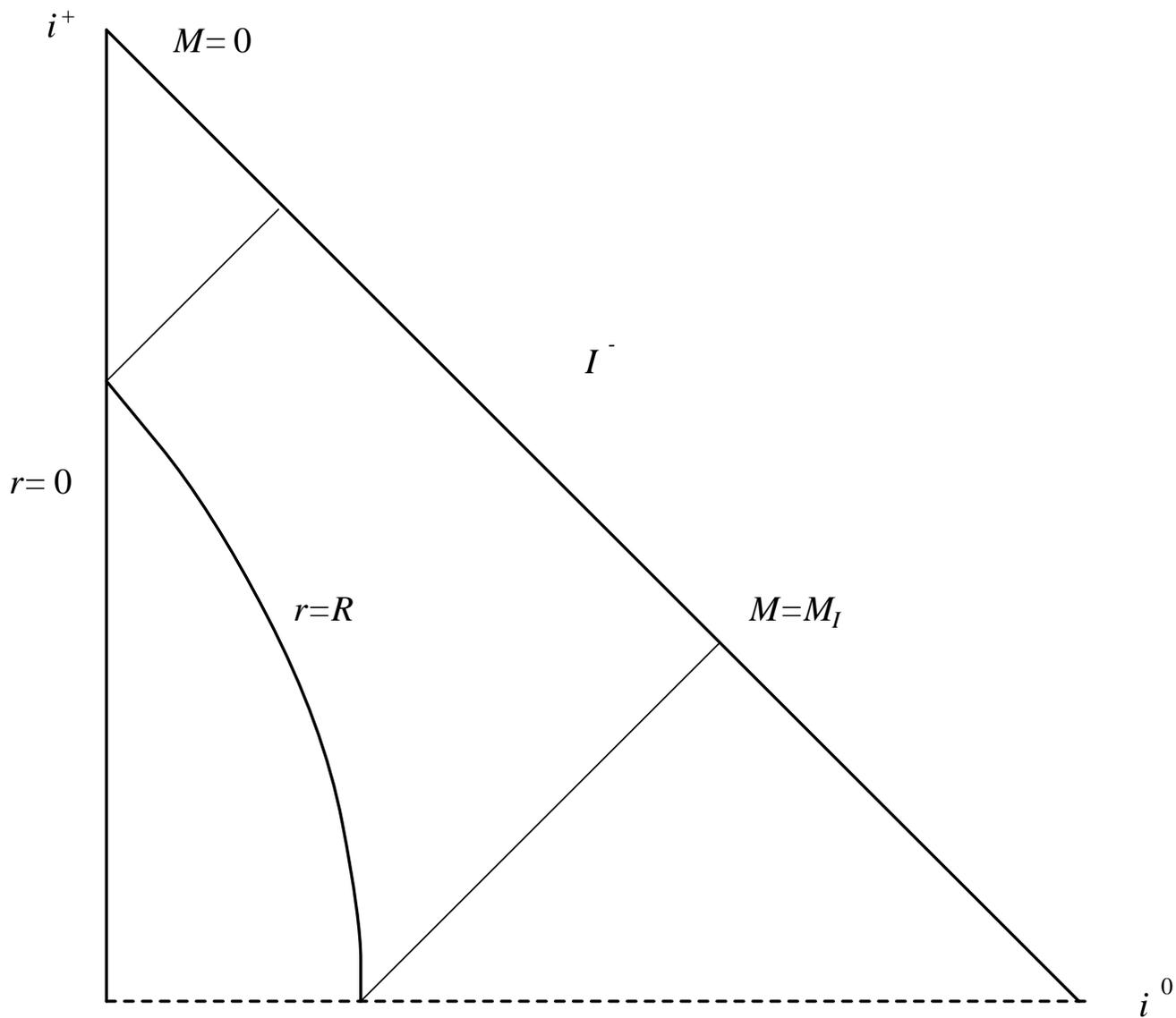
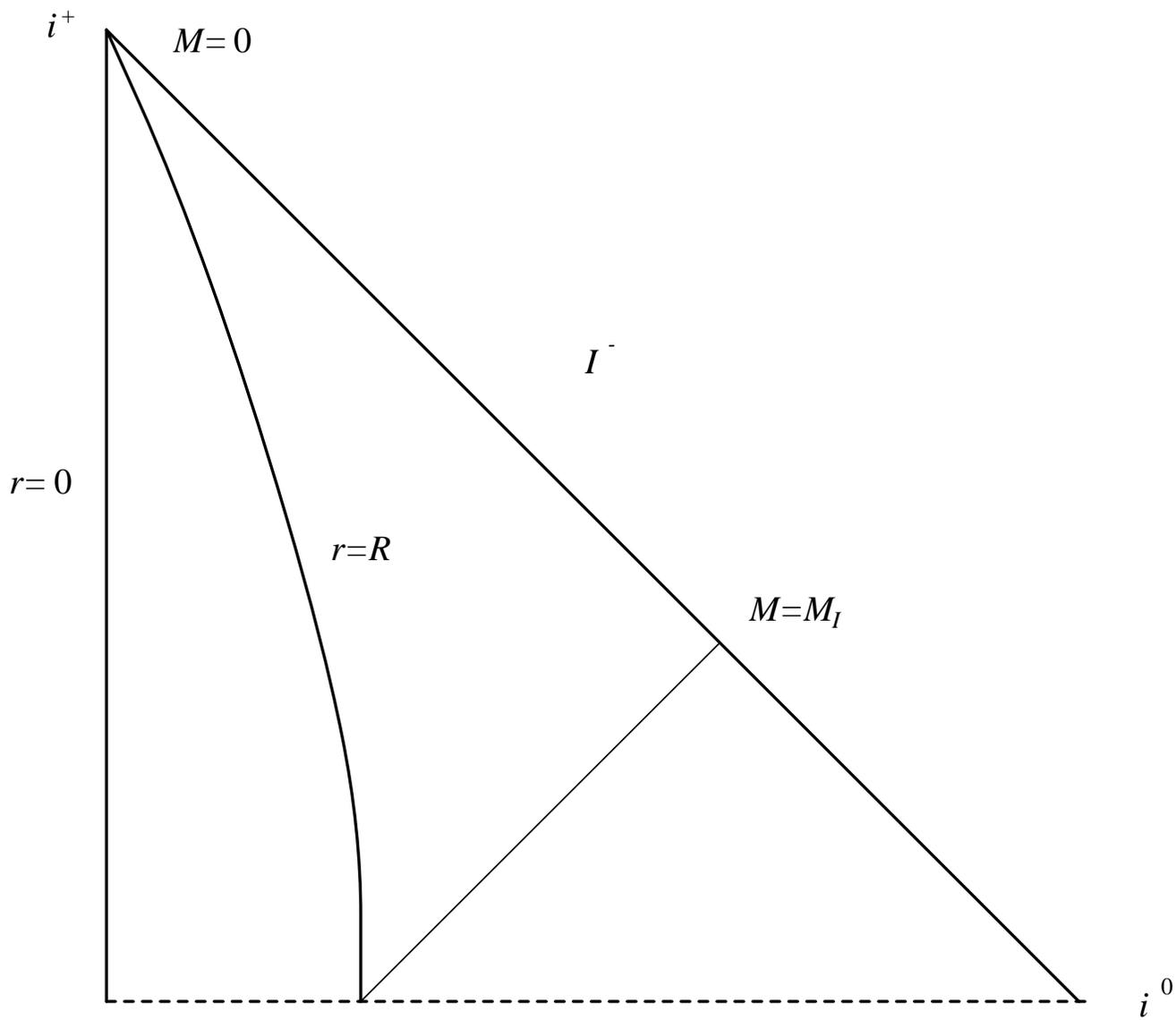
Fig. 2. The Penrose diagram of (a)the supercritical solution,
(b)the critical solution, and (c)the subcritical solution,
where the double lines denote singularities.
In (b) the $r=0$ line above the point of contraction of the shell
should be replaced by double line. 

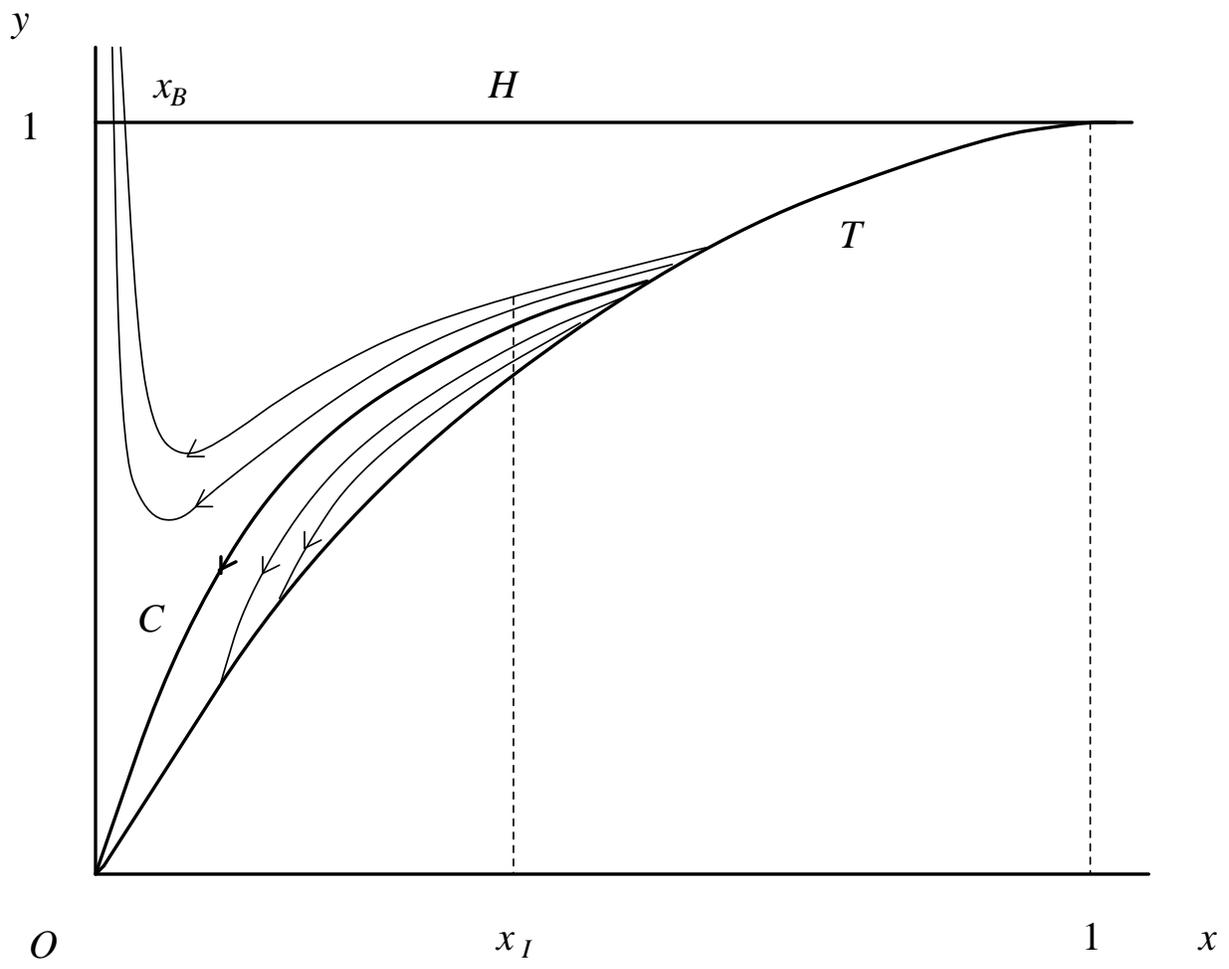
Fig. 3. The evolution in variables $x$ and $y$. 
The line $H$ corresponds to the apparent horizon and
the line $T$ is the set of states with $\dot R=0$. 
Region I is between $H$ and $T$, where we can set initial data.
The thick line $C$ with an arrow denotes the critical solution which
hits the origin.  The thin line with an arrow 
above $C$ is a supercritical solution which crosses $H$.
The coordinate $x_B$ of the point intersecting $H$ gives 
the mass of the formed black hole.
The thin line with an arrow below $C$ is a subcritical solution.

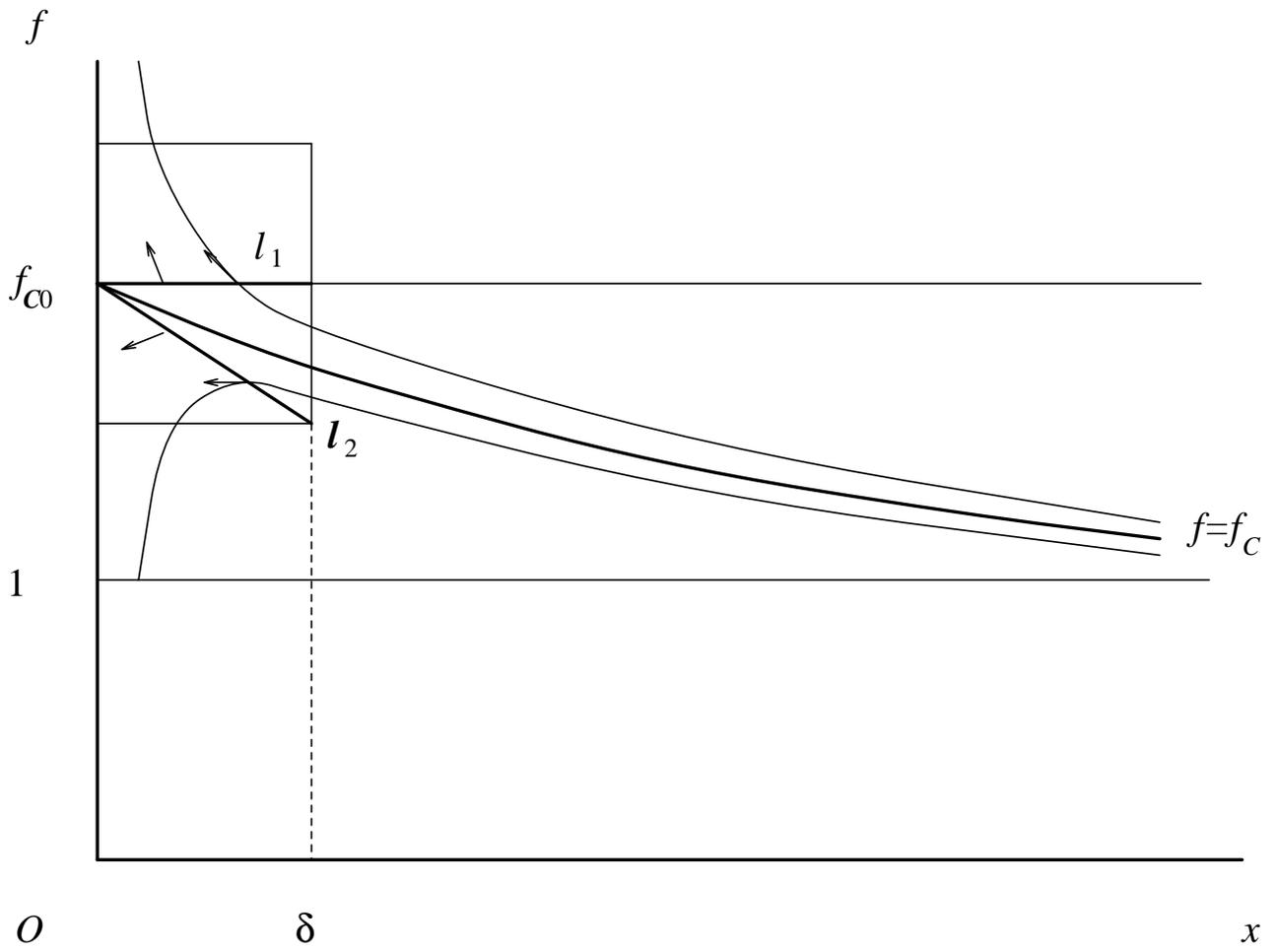
Fig. 4. The behavior of the solutions in variables $x$ and $f$.
The curves are the solutions. The arrows on $l_1$ and $l_2$ indicate 
gradients of the solutions at those points. There must exist some 
solution $f=f_C$ which has a finite limit
$\lim_{x\rightarrow0}f=f_{C0}$. 

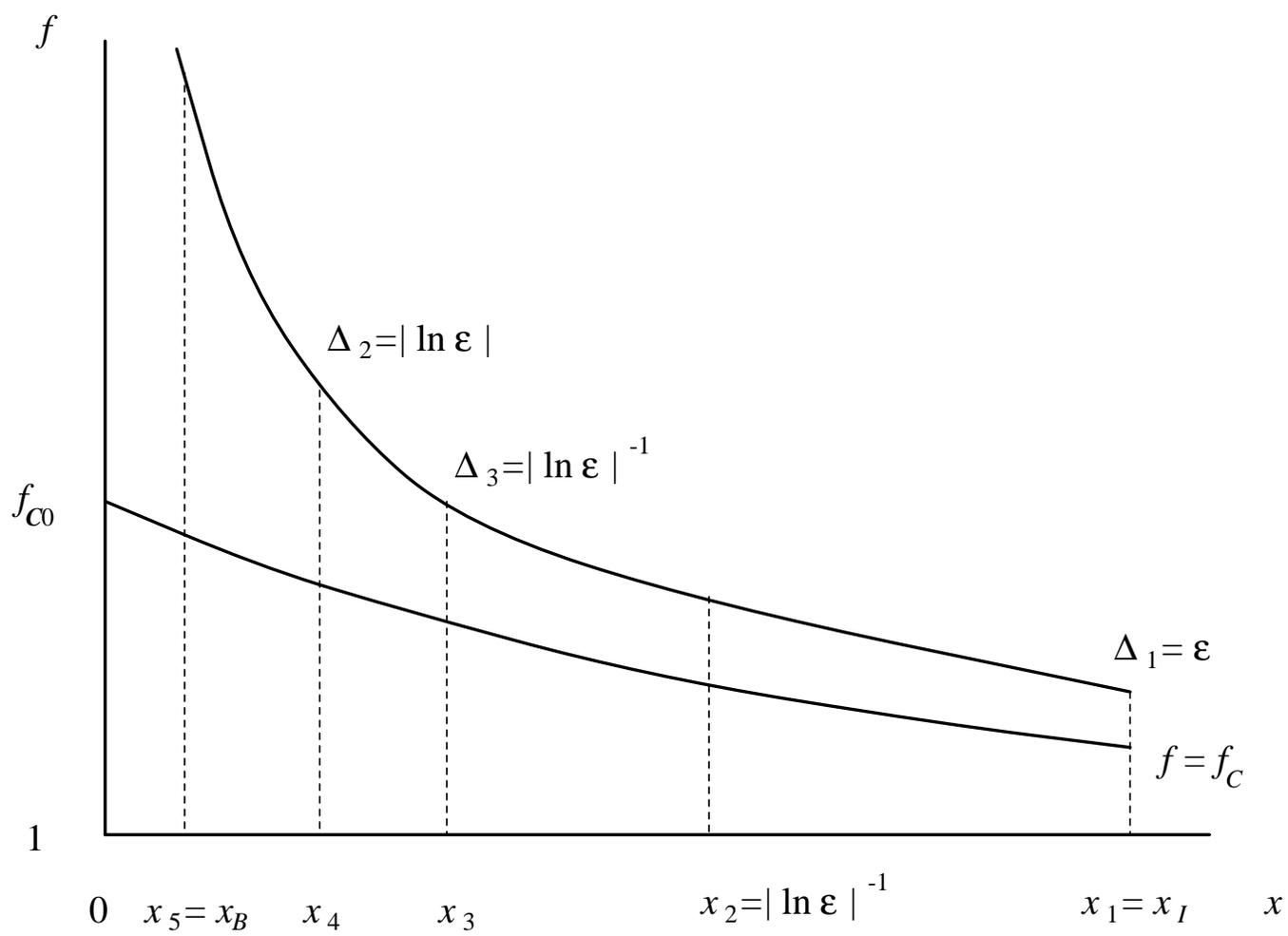
Fig. 5. A figure which illustrates the procedure of finding the exact
value of the critical exponent $\beta$.


\begin{thebibliography}{10}

\bibitem{Cho93}
M.~W. Choptuik,
\newblock {\em Phys. Rev. Lett.} {\bf 70} (1993) 9.

\bibitem{AbEv93}
A.~M. Abrahams and C.~R. Evans,
\newblock {\em Phys. Rev. Lett.} {\bf 70} (1993) 2980.

\bibitem{EvCo94}
C.~R. Evans and J.~S. Coleman,
\newblock {\em Phys. Rev. Lett.} {\bf 72} (1994) 1782.

\bibitem{ONT94}
Y.~Oshiro, K.~Nakamura, and A.~Tomimatsu,
\newblock {\em Prog. Theor. Phys.} {\bf 91} (1994) 1265.

\bibitem{HMN94}
V.~Husain, E.~A. Martinez, and D.~Nunez,
\newblock Exact solution for scalar field collapse, 1994,
\newblock gr-qc/9402021.

\bibitem{Bra94}
P.~R. Brady,
\newblock Does scalar field collapse produce `zero mass' black holes?, 1994,
\newblock gr-qc/9402023.

\bibitem{Isr66}
W.~Israel,
\newblock {\em Nuovo Cim.} {\bf 44B} (1966) 1.

\bibitem{MSY94}
T.~Mishima, H.~Suzuki, and N.~Yoshino,
\newblock Creation of a child universe in Vaidya spacetimes with out-going
  negative energy radiation,
\newblock in {\em Proceeding of the Third Workshop on General Relativity and
  Gravitaion}, edited by K.~Maeda~{\em et. al.}, pages 375--381, Tokyo, 1994.

\bibitem{BGG87}
S.~K. Blau, E.~Guendelman, and A.~H. Guth,
\newblock {\em Phys. Rev.} {\bf D35} (1987) 1747.

\bibitem{His81}
W.~Hiscock,
\newblock {\em Phys. Rev.} {\bf D23} (1981) 2813; {\em ibid.}, 2823.

\bibitem{OKNT93}
Y.~Oshiro, S.~Konno, K.~Nakamura, and A.~Tomimatsu,
\newblock {\em Prog. Theor. Phys.} {\bf 89} (1993) 77.

\end{thebibliography}
\end{document}